\documentclass[]{aa}
\usepackage{graphicx}

\newcommand{\eq}[1]{Eq.~(\ref{#1})}
\newcommand{\fg}[1]{Fig.~\ref{#1}}
\renewcommand{\vec}[1]{\mathbf #1}
\renewcommand{\verb}[1]{\begin{verbatim} #1 \end{verbatim}}
\bibpunct{(}{)}{;}{a}{}{,}
\newcommand{\ca}{\ion{Ca}{II}}
\newcommand{\ha}{H$\alpha$}
\newcommand{\kms}{km~s$^{-1}$}
\begin{document}

\title{Inference of  chromospheric plasma parameters on the Sun}
\subtitle{ Multilayer spectral inversion of strong absorption lines}
\titlerunning{Multilayer spectral inversion}
\authorrunning{Chae et al.}
\author{Jongchul Chae, Maria S. Madjarska, Hannah Kwak, Kyuhyoun Cho}
\institute{Astronomy Program, Department of Physics and Astronomy, Seoul National University, Seoul 08826, Korea}
\abstract
{The solar chromosphere can be observed well through strong absorption lines.
We infer the physical parameters of  chromospheric plasmas  from these lines
  using a multilayer spectral inversion. This is a new technique of spectral inversion.
 We assume that the atmosphere consists of a finite number of layers.
In each layer  the absorption profile   is  constant and the source function varies with optical depth   with a constant gradient. Specifically, we consider a three-layer model of radiative transfer where the lowest layer is identified with the photosphere and  the two upper layers are identified with the chromosphere. The absorption profile in the photosphere is described by a Voigt function, and the profile in the chromosphere by a Gaussian function.  This three-layer model is fully specified by 13 parameters. Four parameters can be fixed  to prescribed  values, and one parameter can be determined from the analysis of a satellite photospheric line. The remaining 8 parameters are determined from a constrained least-squares fitting. We applied the multilayer spectral inversion to  the spectral data of  the H$\alpha$ and the \ion{Ca}{ii} 854.21 nm lines taken in a quiet region  by the Fast Imaging Solar Spectrograph (FISS) of the Goode Solar Telescope (GST).  We find that our model successfully fits most of the observed profiles and produces regular maps of the model parameters.  The combination of the inferred Doppler widths of the two lines  yields reasonable estimates of temperature and nonthermal speed in the chromosphere. We conclude that our multilayer inversion is useful to infer  chromospheric plasma parameters on the Sun.}

\maketitle

\section{Introduction}
Strong absorption lines in the visible and  infrared wavelengths are important spectral windows into the solar chromosphere.
These lines  are observable from the ground and contain useful information of chromospheric plasmas.
The H$\alpha$ line of hydrogen has been the most popular of these windows.
This line is favored because it is  strong, and broad enough for filtergraph observations.
The H$\alpha$ filter images of solar regions display a great  variety of intensity structures \citep{Rutt08,Leen12}.
Even though the image data of the intensity directly provide much useful (mostly morphological) information of the underlying plasma structures, they do not provide estimates of plasma parameters, which are crucial for understanding the physical conditions. The inference of plasma parameters requires the spectral data of the strong absorption lines and a successful spectral inversion.

Spectral inversion is the process of inferring the plasma parameters from the  observed profile of a spectral line.
%
Two types of spectral inversion have been popular in solar observations that assume the constancy of physical parameters. One is the Milne-Eddington inversion, and the other is the cloud model inversion \citep{Beckers1964}.  The Milne-Eddington inversion is based on the assumption that the spectral line is formed in a plasma layer of infinite optical thickness where the absorption profile is constant over optical depth and the source function varies with a constant gradient.   This inversion has been used mainly to model spectral lines formed in the photosphere and to infer the magnetic fields from their Stokes profiles \citep{Unno56,Skum87}.

 The cloud model inversion, on the other hand, assumes that the line is formed in a plasma layer of finite optical depth where the source function as well as the absorption profile is constant over optical depth. This model has been used mostly to infer the physical parameters of cloud-like features lying far above the solar surface \citep{Tzio07}, as was well illustrated in Fig. 1 of \cite{1999A&A...346..322H}.
  A number of variants have been proposed to generalize the original cloud model of \citet{Beckers1964}  by incorporating the varying source function \citep{MeinN1996,1999A&A...346..322H,Tsiropula1999},  the presence of multiple clouds \citep{Gu1996},  the concept of the embedded cloud \citep{Steinitz1977, Chae2014}, etc. Despite these variants, the usage of the cloud model inversion is still limited, and is often hampered by the difficulty of choosing the incident intensity profile. Because the incident intensity below the feature of interest cannot be determined from observations, it has to be assumed to be the same as that in its neighborhood, for instance.

Here we present a multilayer inversion for modeling the spectral profiles of strong absorption lines.  This represents a combined generalization of  the two types of spectral inversion.  A strong line is formed over a wide height range of the atmosphere from the photosphere to the chromosphere. The formation of the line in the photosphere can be modeled by the Milne-Eddington model, and the formation in the chromosphere can be modeled by the cloud model inversion. When the source function is allowed to vary with optical depth, there is no fundamental difference between the two types of inversion.  Thus we expect that the formation of a strong line can be modeled by the radiative transfer across a finite number of layers in each of which  the absorption profile is constant and the source function varies with a constant gradient over optical depth. This is the multilayer inversion we aim to implement.

In this multilayer inversion, parameters other than the source function are kept  constant in each layer, but they  can vary from  layer to layer throughout the solar atmosphere.  In this regard,  the multilayer inversion method is  somewhat similar to  the response-function-based inversion proposed by \citet{1992ApJ...398..375R,1994A&A...283..129R} to infer  height-varying temperature, magnetic field, and line-of-sight velocity  from Stokes profiles.

The multilayer spectral inversion is described in detail in the following section. Its specific version, the three-layer model, is   applied to the spectral data of the H$\alpha$  and  \ion{Ca}{II} 854.21 nm lines taken by the Fast Imaging Solar Spectrograph (FISS) of the Goode Solar Telescope (GST). We measure wavelengths in nm and pm,  and lengths or distances on the Sun   in  km and Mm: 1\,nm= 10$^{-9}$\,m = 10\,\AA,  1\,pm = 10$^{-3}$ nm = $10^{-2}$ \AA\ = 10 m\AA, 1\,km = $10^3$\,m, and 1\,Mm= $10^6$\,m.

\section{Multilayer spectral inversion}

\subsection{Multilayer model of radiative transfer}

We assume that the source function  $S$ is  independent of wavelength over the  spectral line, and is given as a function of height in the atmosphere, which is measured by the optical depth $t_\lambda$ at a wavelength $\lambda$.
If $S$ is known as a function of $t_\lambda$,
the intensity emergent out of the atmosphere $I_\lambda$ is given by the solution of the radiative transfer equation,
\begin{equation}\label{basic}
    I_\lambda = \int_0^{\infty} S (t_\lambda) \exp(-t_\lambda) d t_\lambda  \, .
\end{equation}

The atmosphere consists of the chromospheric layer of finite optical thickness $\tau_0$  at the line center and   the  photospheric layer of infinite optical thickness, therefore we can rewrite \eq{basic} as
\begin{equation}\label{basic_model}
  I_\lambda = I_{\lambda,c} +\exp(-\tau_\lambda) I_{\lambda, p}
\end{equation}
in terms of the chromospheric contribution  $I_{\lambda,c}$ and the photospheric contribution $I_{\lambda,p}$ defined by
\begin{eqnarray*}
   I_{\lambda,c} & \equiv &  \int_0^{\tau_\lambda} S (t_\lambda) \exp(-t_\lambda) d t_\lambda  = L(0, \tau_\lambda) \\
   I_{\lambda, p} & \equiv &  \int_{\tau_\lambda}^\infty S (t_\lambda) \exp(-(t_\lambda- \tau_\lambda)) d t_\lambda  = L(\tau_\lambda, \infty) \, .
\end{eqnarray*}
Here the operator $L(t_1, t_2)$ is defined as
\begin{equation}
  L(t_1, t_2) \equiv    \int_{t_1}^{t_2} S(t) \exp(-(t-t_1)) d t \, .
\end{equation}
We can decompose the chromosphere into $N$ layers of equal optical thickness $\delta_0 = \tau_0/N$ at the line center (as illustrated in \fg{tlm} in the case $N=2$). Then we obtain the expression
\begin{equation}
  I_{\lambda,c} =  \sum_{l=1}^{N} \exp(- t_{\lambda, l-1})  L( t_{\lambda, l-1}, t_{\lambda, l} )
,\end{equation}
where
\begin{equation}
t_{\lambda, l} - t_{\lambda, l-1} = r_{\lambda, l} \delta_0
,\end{equation}
which assumes constancy of  $r_{\lambda} \equiv \chi_\lambda/\chi_0$ over optical depth in each layer, where $\chi_\lambda$ is the absorption coefficient at wavelength $\lambda$ and $\chi_0$ is the absorption coefficient at the central wavelength $\lambda_0$ of the line.
The optical thickness of the chromospheric layer is then given by
\begin{equation}
  \tau_\lambda = t_{\lambda, N}  = \sum_{l=1}^N  r_{\lambda, l} \delta_0 \, .
\end{equation}

We assume that the source function has a constant gradient $dS/dt$ in each layer. Then the integration yields  the expression
\begin{equation} \label{sol1}
  L(t_1, t_2) = S_1 (1 - e^{-\Delta}) + \frac{d S}{dt} (1- e^{-\Delta} - \Delta e^{-\Delta} )
,\end{equation}
 written in terms of the source function at top $S_1$ and $\Delta = t_2 - t_1$.

 In the chromospheric layers,  the optical thickness is finite, therefore the gradient can be written in terms of the source function at the  top $S_{l-1}$, and the one  $S_l$ at the bottom of each layer indexed by $l$, and it follows
 \begin{eqnarray}
  L(t_{\lambda, l-1}, t_{\lambda, l}) & & =   S_{l-1} (1 - e^{-\Delta})  \nonumber \\
   & & + (S_l - S_{l-1}) (1- e^{-\Delta} - \Delta e^{-\Delta} )/\Delta  \,
,\end{eqnarray}
with $\Delta = r_{\lambda, l} \delta_0$.

\begin{figure}
  \centering
  \includegraphics[width=8cm]{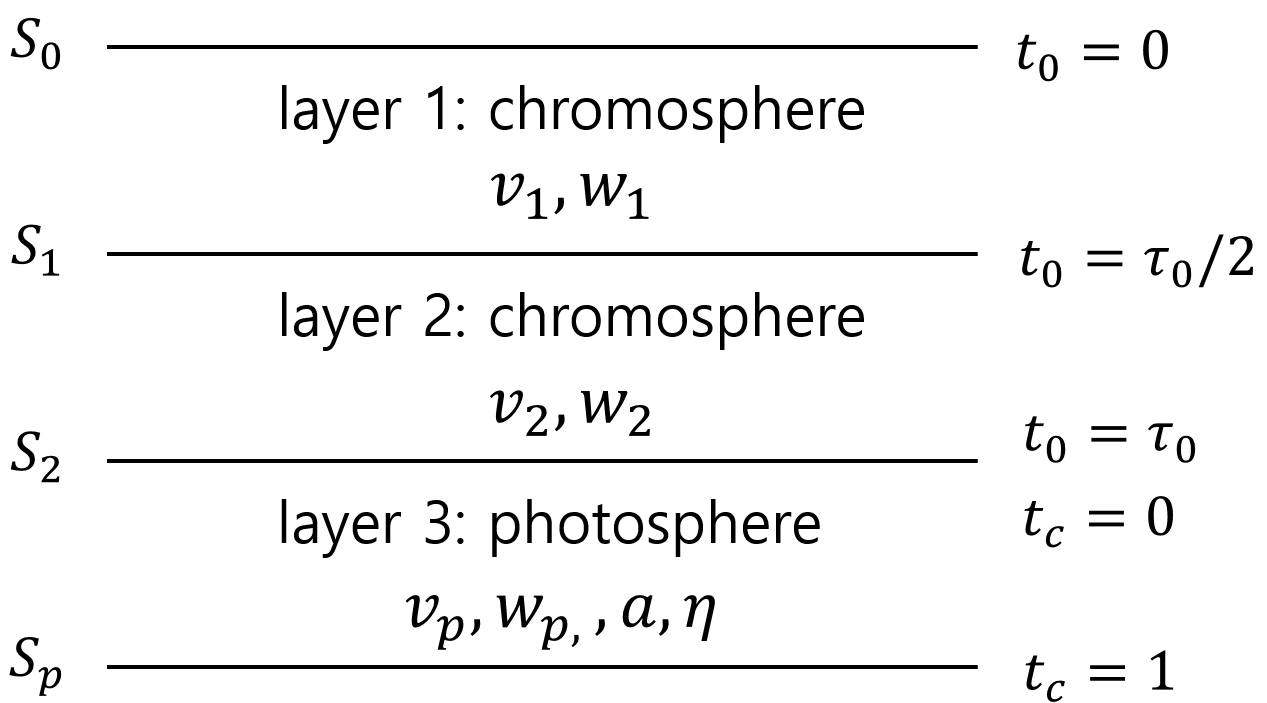}
  \caption{Three-layer model of radiative transfer that consists of two chromospheric layers and one photospheric layer. }\label{tlm}
\end{figure}

The absorption profile is assumed to be constant in each layer. In a chromospheric layer, the values of the line-of-sight velocity $v_{l}$ (defined to be positive in downward motion) and Doppler width $w_l$ are assumed to be constant over optical depth. The damping parameter and  continuum absorption are set to zero. The absorption coefficient in the $l$-th layer is described by the Gaussian function
 \begin{equation}
\chi_{\lambda,l} = \chi_{p, l} \exp (-u_{\lambda, l}^2)
,\end{equation}
with the peak absorption coefficient $\chi_{p, l}$,  and
\begin{equation}
u_{\lambda, l} \equiv  \frac{ \lambda - (1+ v_{l}/c) \lambda_0 }{w_l} .
\end{equation}
The absorption profile at wavelength $\lambda_0$ is given by
\begin{equation}
\chi_{0,l} = \chi_{p, l} \exp (-u_{0, l}^2)
,\end{equation}
with
\begin{equation}
u_{0, l} \equiv -\frac{ v_{l}}{c}  \frac{ \lambda_0 }{w_l} ,
\end{equation}
and we obtain
\begin{equation}
  r_{\lambda, l} = \exp\left(-u_{\lambda, l}^2+u_{0, l}^2 \right)  \, .
\end{equation}

In the photosphere,  $\Delta$ is infinite in \eq{sol1}, and the gradient can be expressed as
\begin{equation}
\frac{d S}{d t_\lambda} = \frac{d S}{d t_c} \frac{dt_c}{d t_\lambda} = (S_p - S_N) \frac{dt_c}{d t_\lambda}
,\end{equation}
where  $t_c$ is the continuum optical depth and $S_p$ is the source function at the level of $t_c=1$.   Thus \eq{sol1} leads to the  expression
 \begin{equation}
  L(\tau_\lambda, \infty) = S_N + (S_p - S_N) \frac{ d t_c}{d t_{\lambda}}
\end{equation}
in the photosphere, where $S_N$ is the source function at the bottom of the lowest layer indexed by $N$.

%



The absorption profile is constant over depth in the photosphere as well.  Here the continuum absorption is not negligible and  the collisional broadening also has to be taken into account.  The ratio of line center-to-continuum absorption $\eta$ as well as the dimensionless damping parameter $a$ is taken to be constant over optical depth.  The absorption coefficient in the photosphere is then given by
\begin{equation}
\chi_{\lambda, p} = \chi_c + \chi_{p, p}  H(u_{\lambda, p}, a)
,\end{equation}
where $\chi_{p,p}$ is the peak absorption coefficient in the photosphere, and $H$ is the Voigt function normalized to satisfy $H(0,a)=1$
with
 \begin{equation}
 u_{\lambda,  p} = \frac{ \lambda -  (1+ v_{p}/c)\lambda_0 }{w_p} \, .
 \end{equation}
 Thus we have
\begin{equation}
  \frac{ d t_c}{d t_\lambda} = \frac{\chi_{c}}{\chi_{\lambda, p}}= \frac{1}{1+ \eta H(u_{\lambda, p}, a)}  \,
,\end{equation}
with the line center-to-continuum opacity ratio $\eta \equiv \chi_{p, p}/\chi_c$.

In this work, we specifically consider the three-layer model of radiative transfer where the chromosphere is assumed to consist of two layers ($N=2$), as illustrated in \fg{tlm}.
We are mainly interested in the chromosphere. Because it covers a wide range of heights over which the physical conditions vary, it is necessary to describe it with a model of at least two layers. The three-layer  model is the simplest model that includes the photospheric layer and characterizes both the height variation of the absorption profile and the nonlinear variation of the source function inside the chromosphere.

 This model is fully specified by a total of 13 parameters.
   The optical thickness of the chromosphere is specified by $\tau_0$. The variation in source function  is specified by the four parameters  $S_0$, $S_1$, $S_2$ , and $S_p$, and the absorption profile is described by $v_p$,  $w_p$, $a$, and $\eta$ in the photosphere,  by $v_2$ and $w_2$ in the lower chromosphere,  and by $v_1$ and $w_1$ in the upper chromosphere.

\begin{table}[t]
  \centering
  \begin{tabular}{|l|cc|}
    \hline \hline
    Parameter & H$\alpha$  & \ion{Ca}{II} 854.2 nm  \\ \hline
    $v_p$      & \ion{Ti}{II} 655.96 nm  &  \ion{Si}{I} 853.62 nm  \\
    $\log \eta$  & 0.51 & 0.43 \\
    $\log w_p$/pm & 0.84 & 0.73 \\
    $\log a$ & 1.10 & 1.32 \\
    $\log \tau_0$ & 1.00 & 1.00 \\
    \hline
  \end{tabular}
  \caption{Fixed parameters and their values. The value of $v_p$ is determined from the specified line. }\label{tbfix}
\end{table}
\begin{table}[t]
  \centering
  \begin{tabular}{|l|rr|}
    \hline \hline
    Parameter & H$\alpha$ \  \  & \ion{Ca}{II} 854.2 \\ \hline
    $\log S_p$      & 0.01$\pm$0.10  & 0.02$\pm$0.10   \\
    $\log S_2$  & -0.25$\pm$0.10 & -0.54$\pm$0.10 \\
    $v_2$/[km s$^{-1}]$  &   0.40$\pm$1.5   &  -0.29$\pm$1.50    \\
    $v_1$/[km s$^{-1}]$   &  -0.50$\pm$1.5    & 0.76$\pm$1.50     \\
      $\log w_2$/[pm] & 1.62$\pm$0.03 & 1.46$\pm$0.05 \\
    $\log w_1$/[pm] & 1.47$\pm$0.05 & 1.26$\pm$0.05 \\
    $\log S_1$ & -0.56$\pm$0.05 & -0.30$\pm$0.05 \\
    $\log S_0$ & -0.80$\pm$0.05 & -0.98$\pm$0.09 \\
    \hline
  \end{tabular}
  \caption{Values of $p_{i,e}\pm \epsilon_i$ of the  free parameters. The values of $S_p$, $S_2$, $S_1$, and $S_0$ are measured in unit of the continuum intensity averaged over the quiet region. }\label{tbfree}
\end{table}

 \begin{figure*}
  \centering
  \includegraphics[width=18cm]{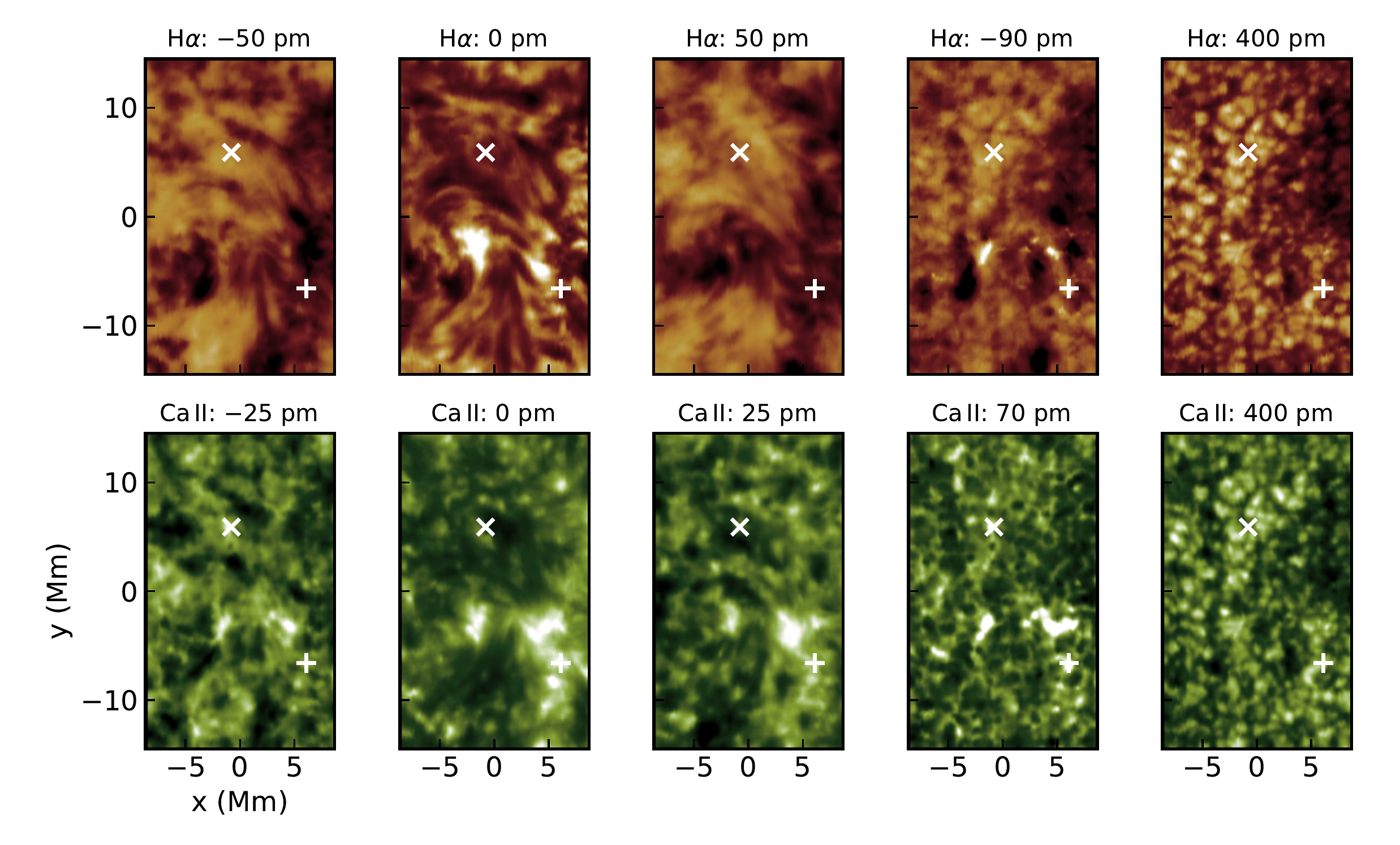}
  \caption{Monochromatic images of a quiet region constructed  at ten selected wavelengths. The two symbols mark the position of a intranetwork feature (x) and a network feature (+) selected for the illustration of the observed line profiles and the  model fitting.
  }\label{raster}
\end{figure*}

\subsection{Model fitting}

We fit each profile of a strong absorption line with the model specified by the independent parameter vector, which is defined as
 \begin{eqnarray}
 \vec p & \equiv & (v_p, \log \eta, \log w_p, \log a,  \log S_p, \log S_2,  \log \tau_0, \nonumber \\
 &  &  v_2, v_1, \log w_2,  \log w_1, \log S_1,  \log S_0 )
 ,\end{eqnarray}
 where each element has a real value.  $\log$ refers to the common logarithm to the base of 10.  The adoption of the logarithmic values at the model parameters automatically  guarantees
 the positivity requirement for  $\eta$, $w_p$, $a$, $S_p$, $S_2$, $\tau_0$,  $w_2$, $w_1$,  $S_1$, and  $S_0$.

For the regularized model  fitting, we employ the technique  of constrained least-squares fitting that minimizes the functional
 \begin{equation}
 H \equiv  \sum_j  \frac{(y_j - x_j (\vec p))^2}{\sigma_y^2}  + \sum_i  \frac{(p_i - p_{i,e})^2}{\epsilon_i^2}   +  \sum_{lm}  \frac{(p_l - p_m)^2}{\epsilon_l^2+\epsilon_m^2}
 .\end{equation}
Here the first sum is the classical $\chi^2$ term where $y_j$ is the data, $\sigma_y$ is the standard noise of data, and  $x_j$ is the model  with parameters $p_i$.  The second sum is the constraint on the individual components $p_i$ , where $p_{i,e}$ and $\epsilon_i$ are the expectation value of $p_i$ and its standard deviation, respectively, that are to be known from the {\em \textup{a priori}} information. The third sum is the similarity  constraint forcing absorption profiles in the two chromospheric layers to be similar to each other as far as the data allow. Specifically, we select $(p_l, p_m) = (v_1, v_2)$,  $(\log w_1,  \log w_2),$ and $(\log S_, \log S_0)$.

Depending on the value of $\epsilon_i$,  the parameter $p_i$ can be categorized as either fixed or free. If $\epsilon_i$ is set to be very small,  $p_i$ is practically fixed to $p_{i, e}$. The value of $v_p$ is inferred from the center of  the  \ion{Ti}{II} 655.958 nm line in the H$\alpha$ band and the \ion{Si}{I} 853.6165 nm line in the \ca\ band. It is thus given as an input, and treated as a fixed parameter in the model fitting.  In addition, we found after several experiments that most line profiles can be fairly well fit with $\log \eta$, $\log w_p$, $a$, and $\tau_0$ being fixed to the values listed in Table~\ref{tbfix}.

 We fix the values of $\log \eta$, $\log w_p$, and $a$, because this reduces the degree of freedom  and facilitates fitting the portions of the line profile that is formed in the photosphere.  We also fix the value of $\tau_0$ because it is not uniquely determined from the fitting. We have examined the performance of the fitting by varying $\tau_0$. As a result, we found that in both  lines, the fitting is fairly good, regardless of $\tau_0$ , because it varies from 1 to 10. The fit becomes worse when  $\tau_0$ becomes larger. Thus we fix $\tau_0$ to 10,  the highest value that can yield a sufficiently good fitting.

There remain eight free parameters  that have to be determined from the fitting itself.  To determine these values, we first apply the fit to a large number of data sets without constraints (by setting $\epsilon_i = \infty$).  Then  all $\vec p$ obtained with the best fit form an ensemble of $\vec p$. The mean of $p_i$ over this ensemble is then identified with $p_{i,e}$.  The value of $\epsilon_i$ is set to the standard deviation of $p_i$ in the ensemble or, if necessary, to a higher value.  The values of $p_{i, e}$ and $\epsilon_i$ are listed in Table~\ref{tbfree}.

All the fits are made with respect to the clean wavelengths, that is, the wavelengths  where the blending by other lines is negligible. We apply the fitting in two stages.   In the first stage,  the far wings of the line profile  (with wavelength offsets larger than 1.2 \AA\  in the H$\alpha$ line and 0.8 \AA\  in the \ca\ line) are fit by the model
\begin{equation}
I_\lambda =  I_{\lambda, p}
,\end{equation}
assuming $\tau_\lambda$ is negligibly small. This first fitting produces the estimates of two parameters $\log S_p$ and $\log S_2$, and the construction of $I_{\lambda, p}$ over the whole line. With $I_{\lambda, p}$ being determined,  we can construct the contrast profile,
\begin{equation}
C_\lambda \equiv \frac{I_\lambda - I_{\lambda, p}}{I_{\lambda,p}} =
 \frac{I_{\lambda, c}}{I_{\lambda, p}} - (1 -\exp(-\tau_\lambda)) .
\end{equation}
This contrast profile is fit by the corresponding model at the clean wavelengths.  This second fitting
 yields the estimates of the other six parameters:  $v_{2}$, $v_{1}$,  $\log w_2$, $\log w_1$, $\log S_1$, and $\log S_0$.
The special condition of  $v_{2}= v_{1}$,  $w_1$, and  $S_2 = S_1 = S_0 \equiv S $
reduces the above equation to that of the classical cloud model.

The goodness of fit is measured by the standard error $\epsilon$  defined by the root mean square of the difference between the observed $C_\lambda$ and the model $C_\lambda$ , where the average is taken over the clean wavelengths where the fitting is applied.


 \begin{figure*}
  \centering
  \includegraphics[width=14cm]{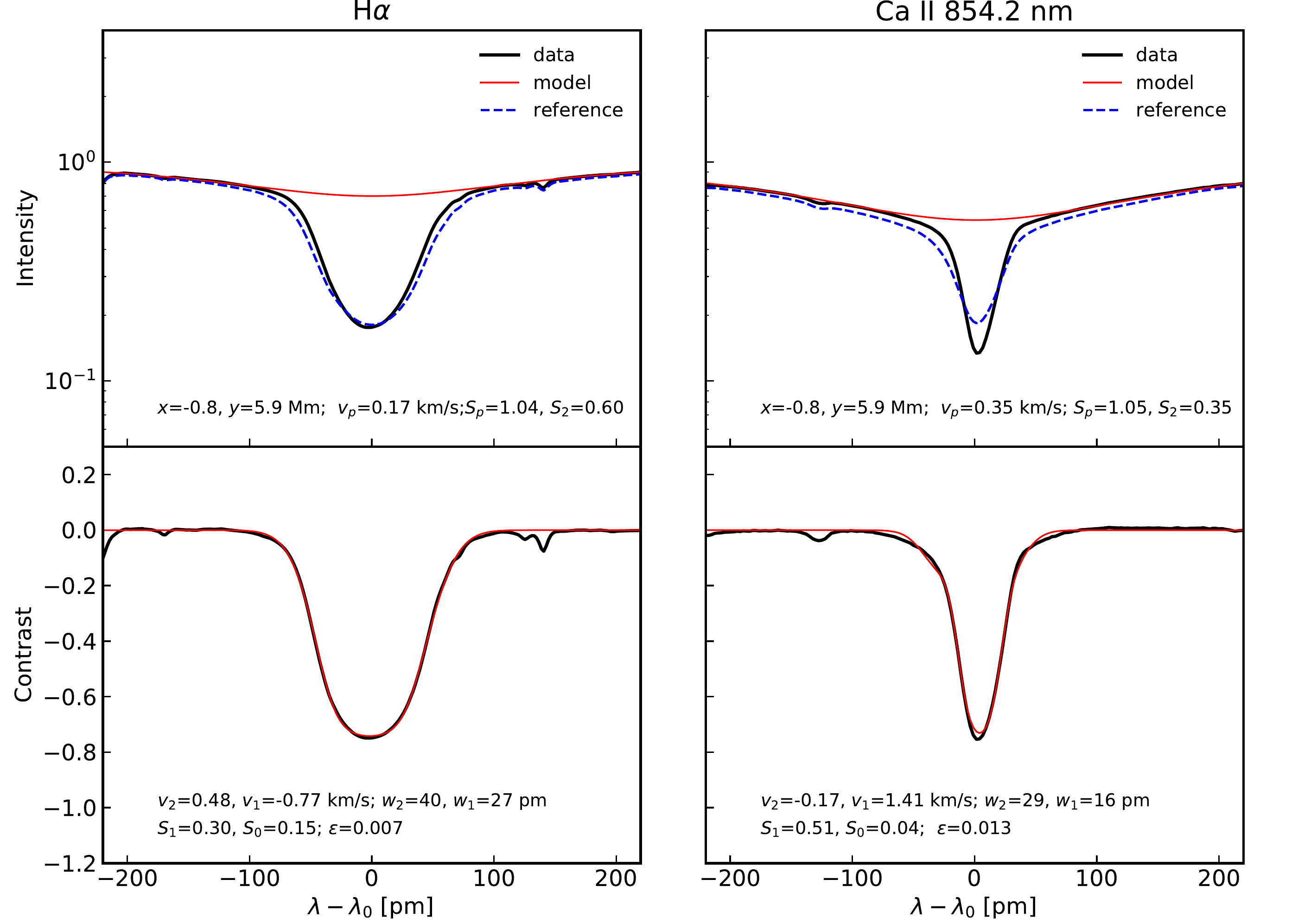}
  \caption{Three-layer model fitting of the line profiles taken from an intranetwork feature marked by the  cross in \fg{raster}. The reference profile is the average of all the profiles over the observed region. }\label{line_in}
\end{figure*}

\section{Data and reduction}
\subsection{Data}

We used the spectral data taken with the FISS of the GST at Big Bear Solar Observatory \citep{Chae2013}. The FISS is a dual-band echelle spectrograph that usually records the H$\alpha$ band and the \ca\ 854.2 nm band simultaneously using two cameras. The cameras record the spectral ranges of 0.97 nm (H$\alpha$) and 1.29 nm (\ca) with a pixel sampling of  1.9 pm and 2.5 pm, respectively. With a slit width of 0.16\arcsec\ , the spectral resolving power ($\lambda/\delta \lambda$) is estimated at 140000 and 130000. We use the PCA-compressed spectral data  where noise is suppressed very well \citep{Chae2013}. The signal-to-noise ratio of the compressed intensity data is 700 (H$\alpha$) and 250 (\ca) at the line centers.
The height of 40\arcsec\ is covered with the sampling of 0.16\arcsec. The two-dimensional  imaging of the FISS is achieved by scanning the slit across the field of view with a step size of 0.16\arcsec.


\fg{raster} presents the monochromatic images  of an observed region constructed from the raster-scan observation of a quiet-Sun region outside active regions. The time of observation was 17:14:06 UT on 2017 June 14.  The field of view is 24\arcsec $\times$  40\arcsec\ or 17.4 Mm $\times$ 29 Mm.
Even though small, the field of view includes both network features and intranetwork features.

\subsection{Reduction}
The average spectral profile of an observed solar region was taken as the reference profile. All the profiles were normalized by the maximum intensity (a proxy of the continuum intensity)  of this reference profile. Each profile was then corrected for stray light in two steps following \cite{Chae2013}. The observed line profile  was corrected for spatial stray light by subtracting 0.027 times the reference profile from it and by then dividing it by 0.973. The  corrected line profile was then corrected for spectral stray light by subtracting 0.065 times the maximum value of the line profile from it and by dividing it by 0.935.

 The reference profile was  also used to calibrate the wavelength precisely.  In the H$\alpha$ band, the centers of the H$\alpha$ line and the \ion{Ti}{II} 655.9580 nm line  in the reference profile were determined in pixel units and were used to convert the wavelength pixels into physical wavelength values. In the \ca\  band, the \ca\ line itself could not be used as a reference because the  center of this line averaged over a solar region is known to be offset from its laboratory wavelength. Instead, we used the pair of the \ion{Fe}{I} 853.80152 nm line and the \ion{Si}{I} 853.6165 nm line for the wavelength calibration. Thus the Doppler shift in each band was  measured with respect to  the average photosphere of the observed region.

Before applying the model fitting, we corrected each observed line profile for the slightly nonuniform pattern that may exist and may cause the asymmetry between the far blue wing and the far red wing.  We fit the intensities at the far red and blue wings $>$ 0.4 nm  by a first-order polynomial. This fit, after normalizing by its mean value, was used as the estimate of the nonuniform pattern.  The profile corrected for this pattern then had the required symmetry.

 The observed spectral profiles of the \ha\  and the \ca\ lines are contaminated
 by weak spectral lines in the same band, some of which are solar lines and others are terrestrial lines (mostly H$_2$O lines). Fortunately, the terrestrial lines in our data are very weak, because the observatory is located in a dry land. The model fitting described below uses only the spectral data that are less contaminated by these satellite lines.

\begin{figure*}
  \centering
  \includegraphics[width=14cm]{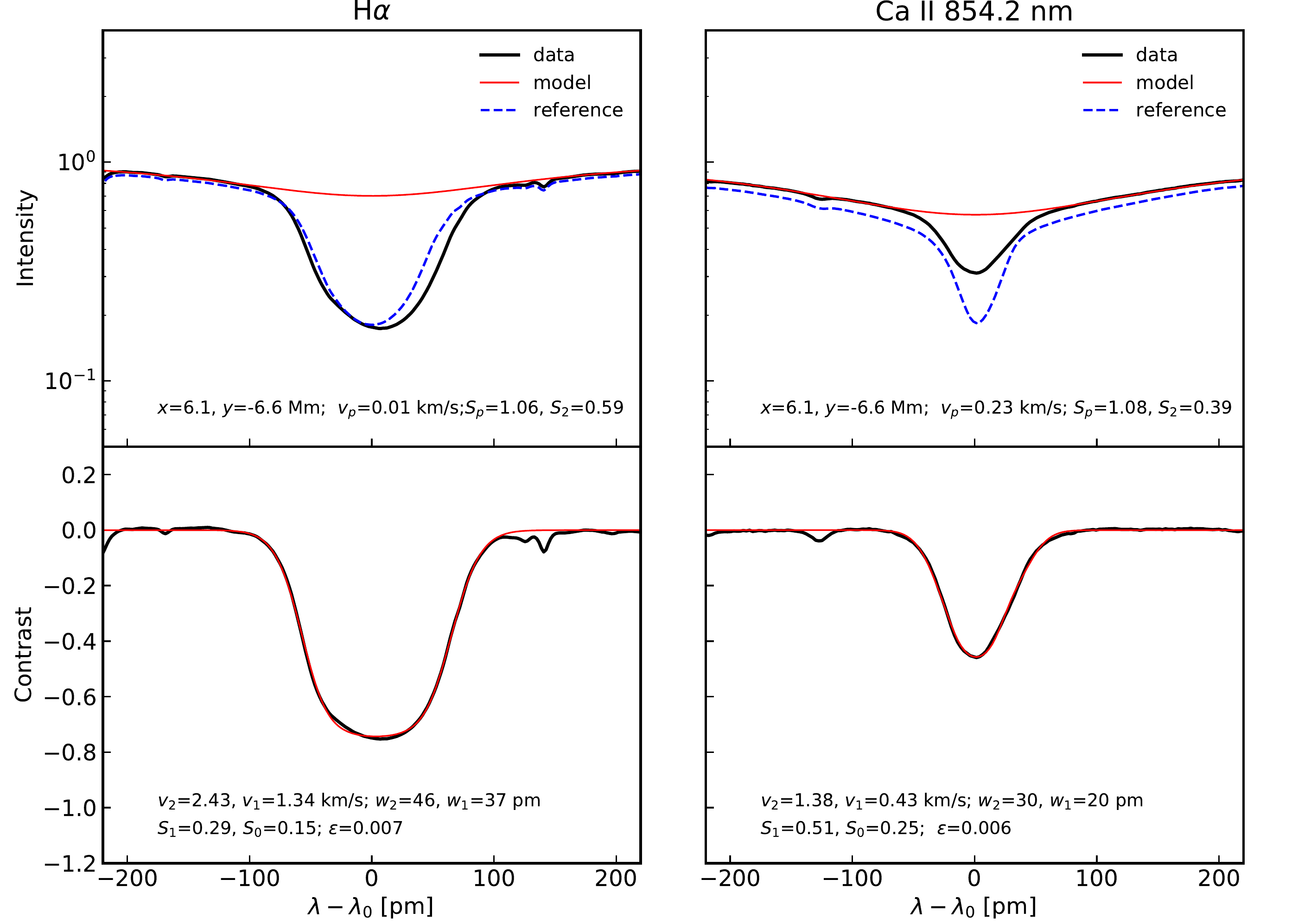}
  \caption{Three-layer model fitting of the line profiles taken from a network feature marked by a plus in \fg{raster}. }\label{line_nw}
\end{figure*}

\begin{figure*}[t]
  \centering
  \includegraphics[width=18cm]{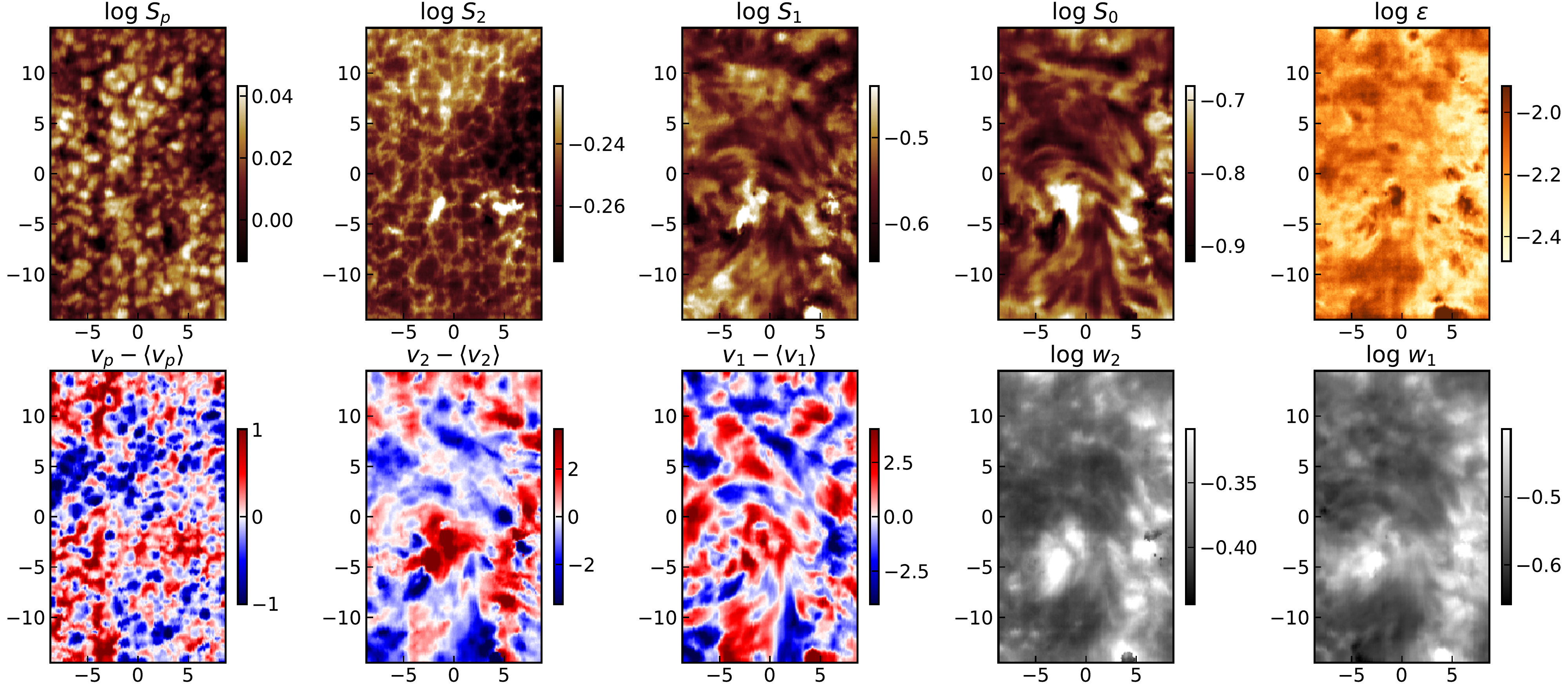}
  \caption{Maps of the H$\alpha$ line parameters. The spatial averages have been subtracted in the  Doppler velocity maps.  }\label{parha}
\end{figure*}

\begin{figure*}
  \centering
  \includegraphics[width=18cm]{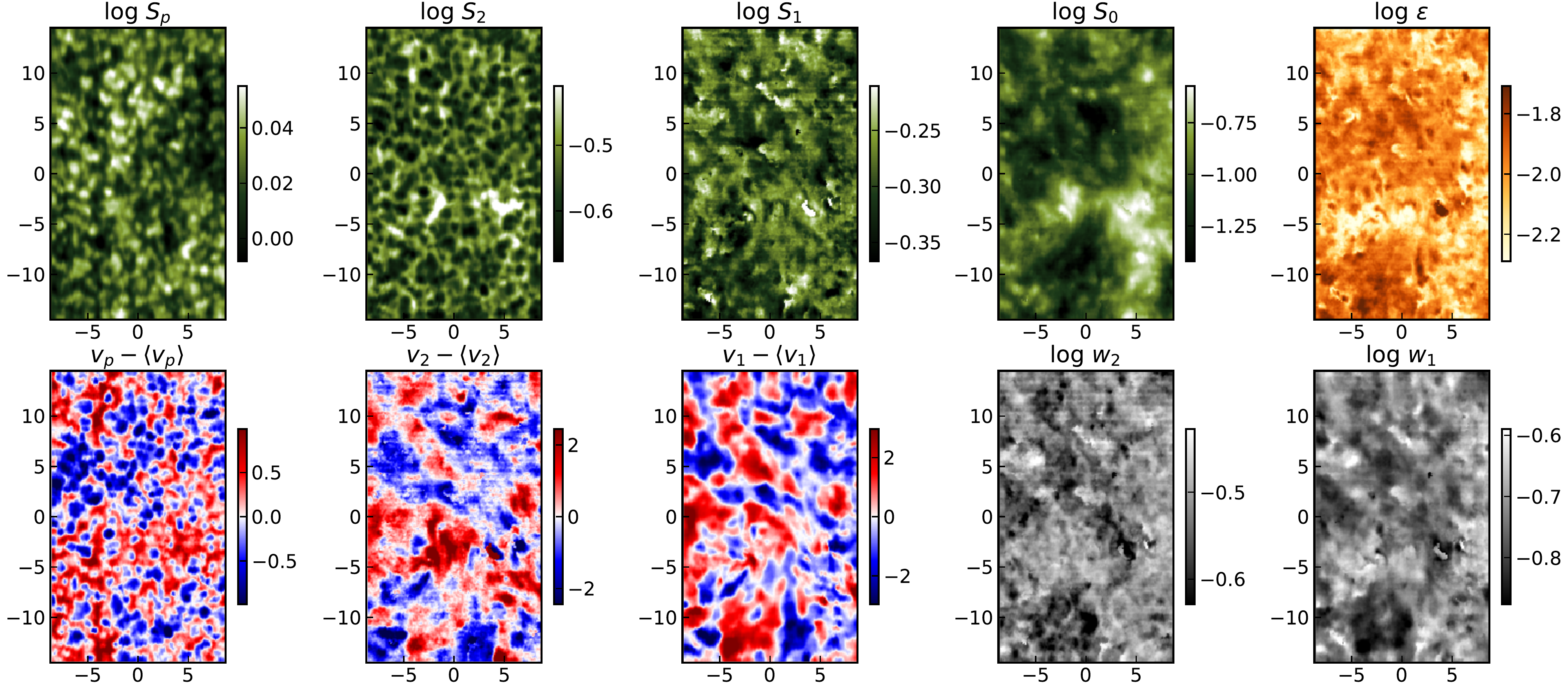}
  \caption{Maps of the  \ca\ line parameters.  The spatial averages have been subtracted in the  Doppler velocity maps.  }\label{parca}
\end{figure*}

\section{Results}

\subsection{Model fitting of line profiles}
We applied the multilayer fitting to the profiles of the \ha\  and \ca\ 854.2 nm lines.   It took 0.018 s  on average to fit  each \ha\ line profile, and  0.046 s to fit each \ca\ line profile
when we used the Python 3.7 software installed on a laptop computer with a 1.80 GHz CPU and Windows 10 operating system.

 We found that the multilayer model fitting is reasonably  good in both lines.
The fitting of the line profiles of an intranetwork (IN) feature is illustrated in \fg{line_in}, and that of a network (NW) feature in \fg{line_nw}.  The standard error of the fitting is $\epsilon=0.007$ (IN)  and 0.007 (NW) in the H$\alpha$ line,  and 0.013 (IN) and 0.006 (NW) in the \ca\ line.
The two figures show that the NW feature is different from the IN feature in the spectral characteristics.  The NW feature has a broader profile of the H$\alpha$ line and a higher core intensity of the \ca\ line than the IN feature. For this reason, the model fitting produces  a larger $w_1$ of the \ha\ line and a higher  $S_0$ of the \ca\ line in the NW feature than in the IN feature.

\subsection{Spatial distribution of the model parameters}
We find from \fg{parha} that the parameter maps of the \ha\ line are very regular.  There are few noticeable irregularities.   The map of $\epsilon$ indicates that the fitting is better than  $\epsilon=0.01$
in most of the spatial pixels in the field of view.  This shows the effectivity of our spectral inversion.   The model parameters can be used  for a reasonable derivation of chromospheric plasma parameters.

The maps of the source function provide information on  temperature or radiation field  at the different atmospheric levels.
The map of $S_p$ is very similar to  the near-continuum intensity image at 0.4 nm off the line
(see \fg{raster}) and corresponds to the spatial distribution of the temperature at the continuum-forming level of the photosphere.  The map of $S_0$ is practically a reproduction of the core intensity image of the line (\fg{raster}).
This is expected  because it is well known that the core of a very strong line is formed
  at the outer part of the atmosphere, and its intensity  is expected to be very close to the source function at the formation level.
The maps of $S_2$ and $S_1$ show the structure of the source function at the top level of the photosphere and at the lower level of the chromosphere, respectively.  The map of $S_2$ displays the inverse convective pattern consisting of dark cells and bright lanes, and the map of $S_1$ is similar to that of $S_0$.

 Our spectral inversion produces the Doppler velocity maps at the three atmospheric levels (\fg{parha}).  These are very useful probes of the atmospheric dynamics. The map of $v_p$ shows the velocity pattern associated with granulation and photospheric oscillations, while
 the maps of $v_2$ and $v_1$ mostly show the velocity pattern   linked to jet-like features
 and chromospheric oscillations. The similarity of $v_2$ and $v_1$ (with the Pearson correlation of 0.56)  suggests that the difference in the oscillation
 phase between the lower chromosphere and the upper chromosphere is not large, implying that the wavelength of the associated waves may be longer than the height difference between the two layers.

The maps of either $w_2$ or $w_1$  of the \ha\ line  (\fg{parha}) provide a convenient way of distinguishing between the network regions and the internetwork regions. The network regions have high values of Doppler width and  the internetwork regions have low values. The map of $w_1$ practically corresponds to the temperature map in the upper atmosphere because  the hydrogen atom is light and its thermal speed dominates the Doppler broadening (compare with \fg{txi}).

\fg{parca} shows the maps of the \ca\ model parameters.  Because the maps of $S_p$, $S_2$ , and $S_0$  are  very regular, they contain information on the spatial distribution of the physical parameters at the corresponding atmospheric level. The map of $S_0$ is very similar to the core intensity image (\fg{raster}). It is  similar to the map of the \ha\ $w_1$ in \fg{parha}. On the other hand, the map of $S_1$ is not regular in that it contains a number of discontinuities at different spatial scales.

The maps of $v_p$, $v_2$, and $v_1$ of the \ca\ band look very similar to those of the \ha\ line.  It is expected that  the \ca\ band $v_p$ and the \ha\ band $v_p$ have very similar patterns (with the Pearson correlation of 0.88) because they were derived from two weak lines that formed in the photosphere. The similarities of  the \ca\ $v_2$ and the \ha\ $v_2$ (correlation=0.72)  and  that of  the \ca\ $v_1$ and the \ha\ $v_1$  (correlation =0.76) are more interesting.  This supports the notion that the formation layer of the \ca\ 854.2 nm line significantly overlaps that of  the \ha\ line.

The maps of the \ca\ $w_2$ and $w_1$ are more complicated than those of the \ha\ $w_2$ and $w_1$.  There is a tendency for $w_2$ and $w_1$  to be larger in the NW regions than in the IN regions. This tendency, however, is not as strong as in the \ha\ line. We find that numerous small patches of enhanced value of $w_1$ are found to be scattered in the IN regions as well as in the NW regions.

We can estimate the random fitting errors by assuming that the  physical conditions and systematic errors vary very smoothly with position so that their second-order spatial derivatives are approximately zero. Then we  assume that the  nonzero second-order finite difference of three neighboring points is contributed by the random errors.  We applied this second-order derivative method to the variation in parameters along the slit direction at each slit position, which provides the estimates of the random error in each parameter. By repeating this process at a number of slit positions and by taking the average, we obtained the representative random errors of the mode parameters as listed in Table~\ref{tberror}.


\begin{table}[t]
  \centering
  \begin{tabular}{|l|rr|}
    \hline \hline
    Parameter & H$\alpha$ \  \  & \ion{Ca}{II} 854.2 \\ \hline
    $S_p$  &  0.0017  & 0.0012   \\
    $S_2$  &  0.0012 &  0.0021 \\
    $S_1$  &  0.0023 &  0.0071 \\
    $S_0$  &  0.0010 &  0.0019 \\
    $v_p$/(km s$^{-1})$  &   0.03   &   0.03 \\
    $v_2$/(km s$^{-1})$  &   0.08   &  0.12    \\
    $v_1$/(km s$^{-1})$   &  0.06   &  0.04    \\
    $w_2$/pm            & 0.13  &  0.38 \\
    $w_1$/pm            & 0.13  &  0.26 \\
     \hline
  \end{tabular}

  \

  \caption{Representative random errors  of the model parameters.  }\label{tberror}
\end{table}

\begin{figure}
  \centering
  \includegraphics[width=8cm]{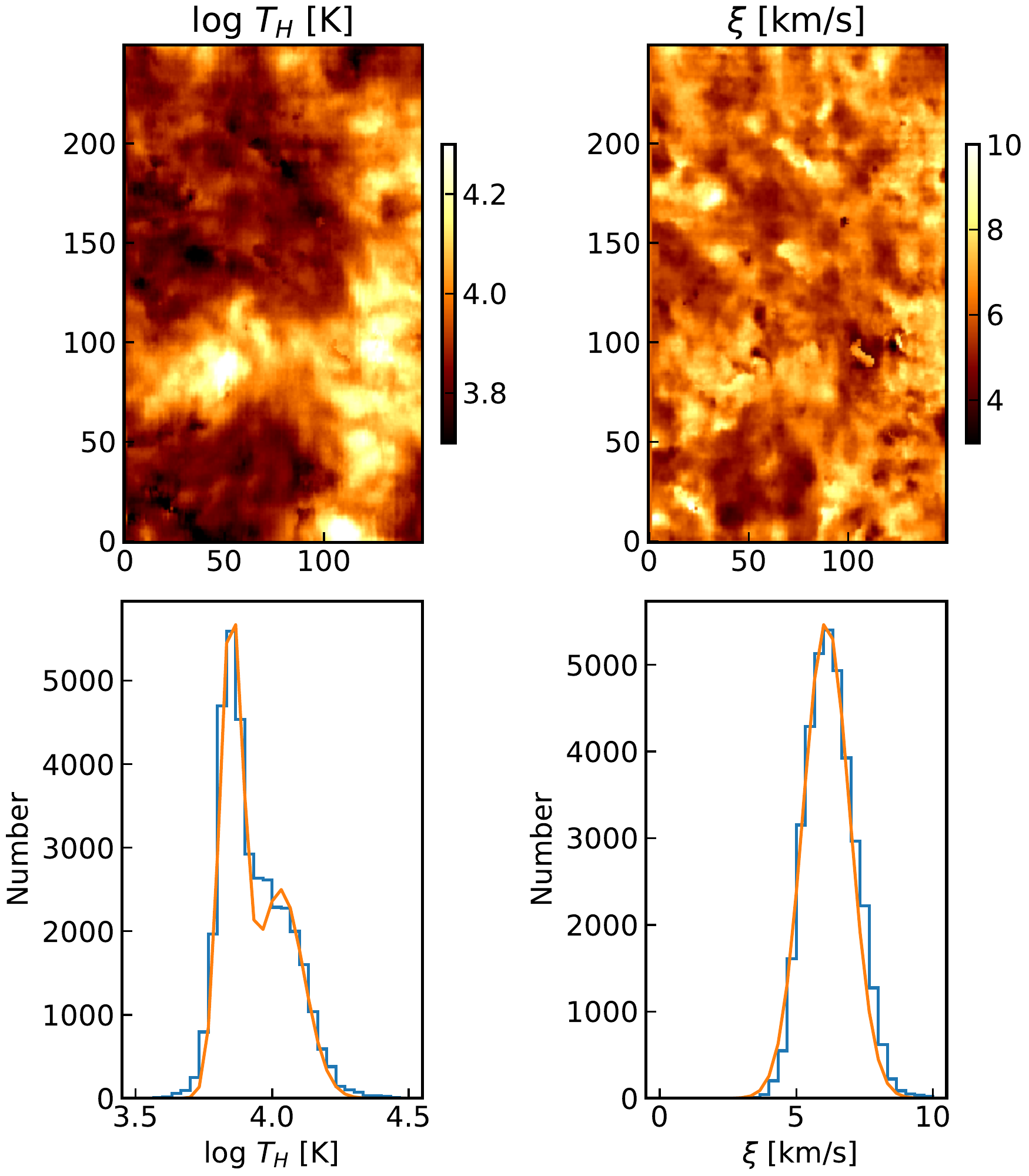}
  \caption{
  Top: Spatial distributions of $T_{\rm H}$ and $\xi$. Bottom: Number distributions of $T_{\rm H}$ and $\xi$. }\label{txi}
\end{figure}

\subsection{Inferring temperature and nonthermal speed}
It is possible to separately determine the hydrogen temperature $T_{\rm H}$ and  the nonthermal speed (or microturbulence speed) $\xi$ of the upper chromosphere by combining the values of $w_1$ of the two lines. By denoting $w_1$ of the \ha\ line by $w_{\rm H}$ and that of the \ca\ line by $w_{\rm Ca}$, we can derive the expressions
\begin{eqnarray}
  T_{\rm H} &=&  8100 \; \hbox{K} \,  \left[ \frac{w_{\rm H}}{\rm 0.025 nm} \right]^2 \left( 1 - 0.59 \left[ \frac{w_{\rm Ca}}{w_{\rm H}}\right]^2 \right) \\
  \xi &=& 5.40 \; \hbox{km\,s}^{-1} \,   \frac{w_{\rm Ca}}{\rm 0.015 nm}   \left( 1 - 0.042 \left[ \frac{w_{H}}{w_{\rm Ca}}\right]^2 \right)^{1/2}
,\end{eqnarray}
which indicate that $T_{\rm H}$ is mostly determined by $w_{\rm H}$ because of the light mass of the  hydrogen atom and $\xi$, mostly by $w_{\rm Ca}$ because of the high mass of a \ca\ ion.

In the IN feature of \fg{line_in}, we have $w_{\rm H}=0.025$ nm and $w_{\rm Ca}=0.015$ nm, which leads to the estimates $T=6500$ K and $\xi=5.1$ \kms.   In the NW feature of \fg{line_nw},    we have $w_{\rm H}=0.036$ nm  and $w_{\rm Ca}=0.023$ nm in the \ca\ line, which leads to the estimates $T=13000$ K and $\xi=7.8$ \kms.

\fg{txi} shows the  spatial distribution of $T_{\rm H}$ and $\xi$ determined from the \ha\ line $w_1$ in \fg{parha} and the \ca\ line $w_1$ in \fg{parca}.  Tthe map of $T_{\rm H}$ is clearly very similar to the \ha\ line $w_1$, and that of $\xi$, to the \ca\ line $w_1$.   Making use of the second derivative method described  above, we estimated the errors of $T_{\rm H}$ and $\xi$ at 140 K and 0.1 \kms.

 The number distribution of $T_{\rm H}$ is approximately a double Gaussian. The peak around 7000 K represents the typical temperature of the upper chromosphere in the IN regions and the other peak at 11000 K, that in the NW regions. In contrast, the number distribution of $\xi$ is singly peaked at 6.1 \kms,  which represents the typical $\xi$ of the upper chromosphere, regardless of the specific region.

\begin{figure}
  \centering
  \includegraphics[width=8cm]{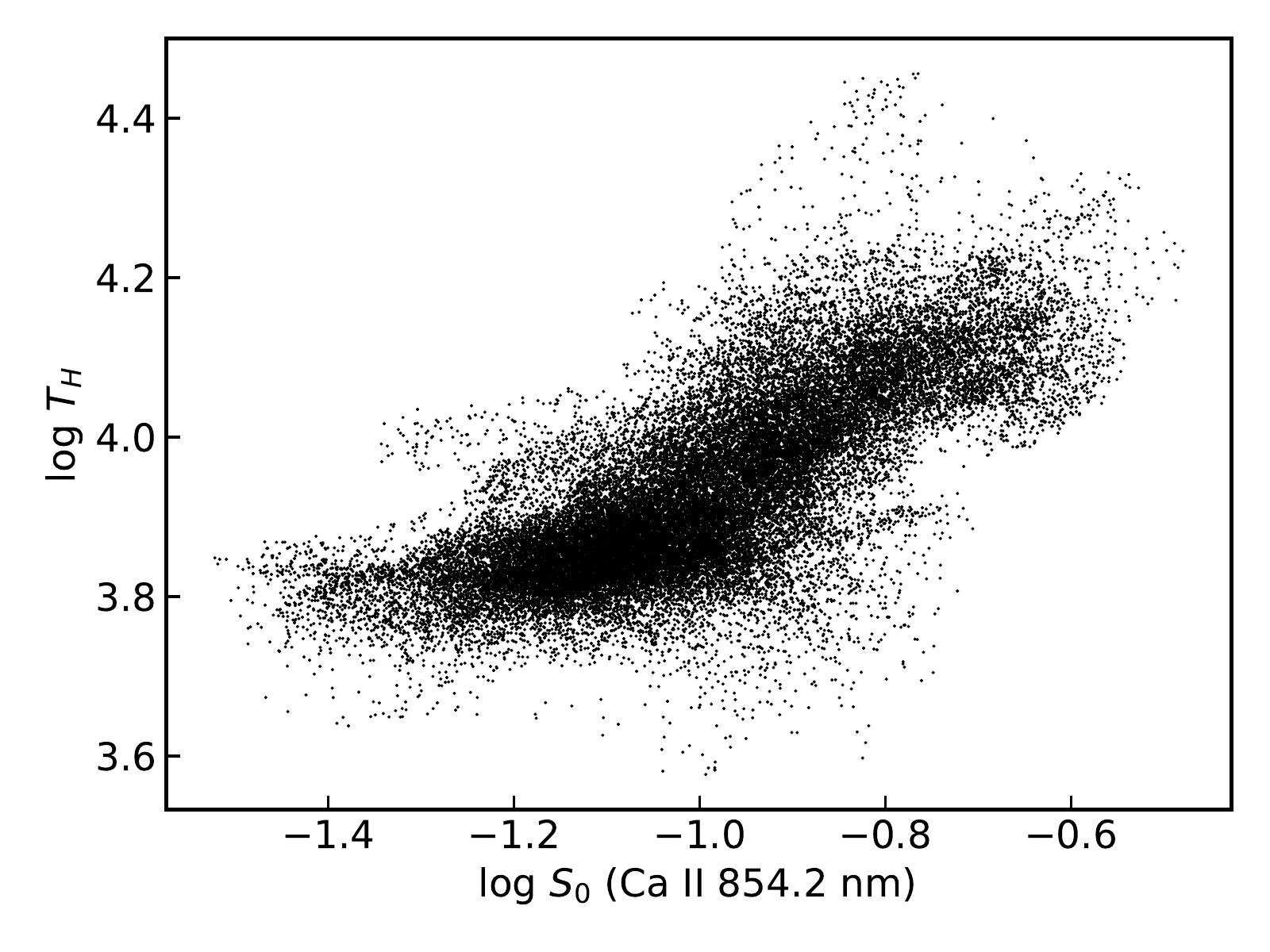}
  \caption{Scatter plot of $T_{\rm H}$ vs. the \ca\ line $S_0$.    }\label{sfT}
\end{figure}
\begin{figure}
  \centering
  \includegraphics[width=9cm]{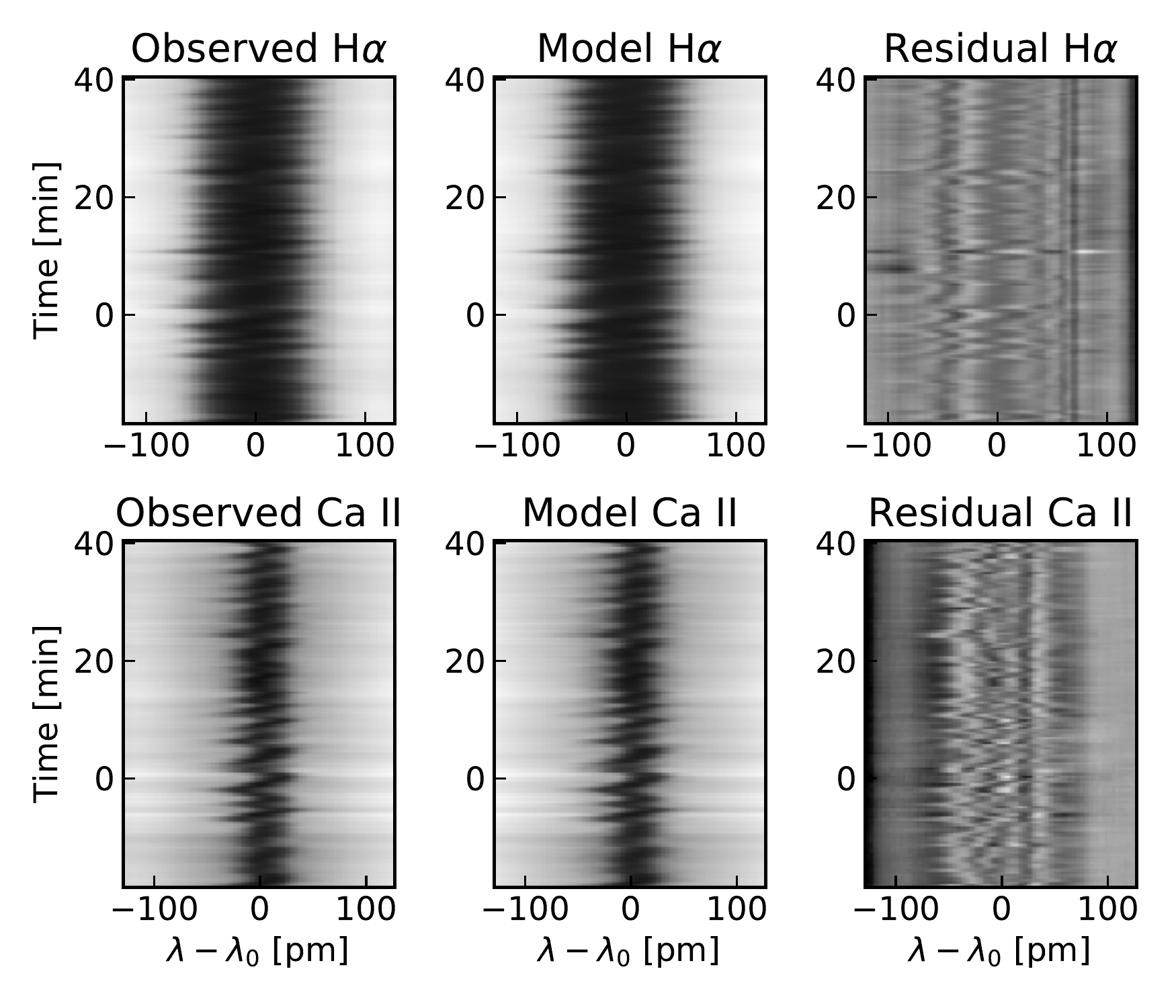}
  \caption{Wavelength-time maps of the \ha\ and \ca\ intensity stacked over time -- observation (\emph{left}) and model (\emph{middle}), and  maps of residual intensity normalized   by the continuum intensity (\emph{right}). The grey scale of the normalized residual intensity spans from -0.03 to 0.03.  }\label{wlt}
\end{figure}

\begin{figure*}
  \centering
  \includegraphics[width=14cm]{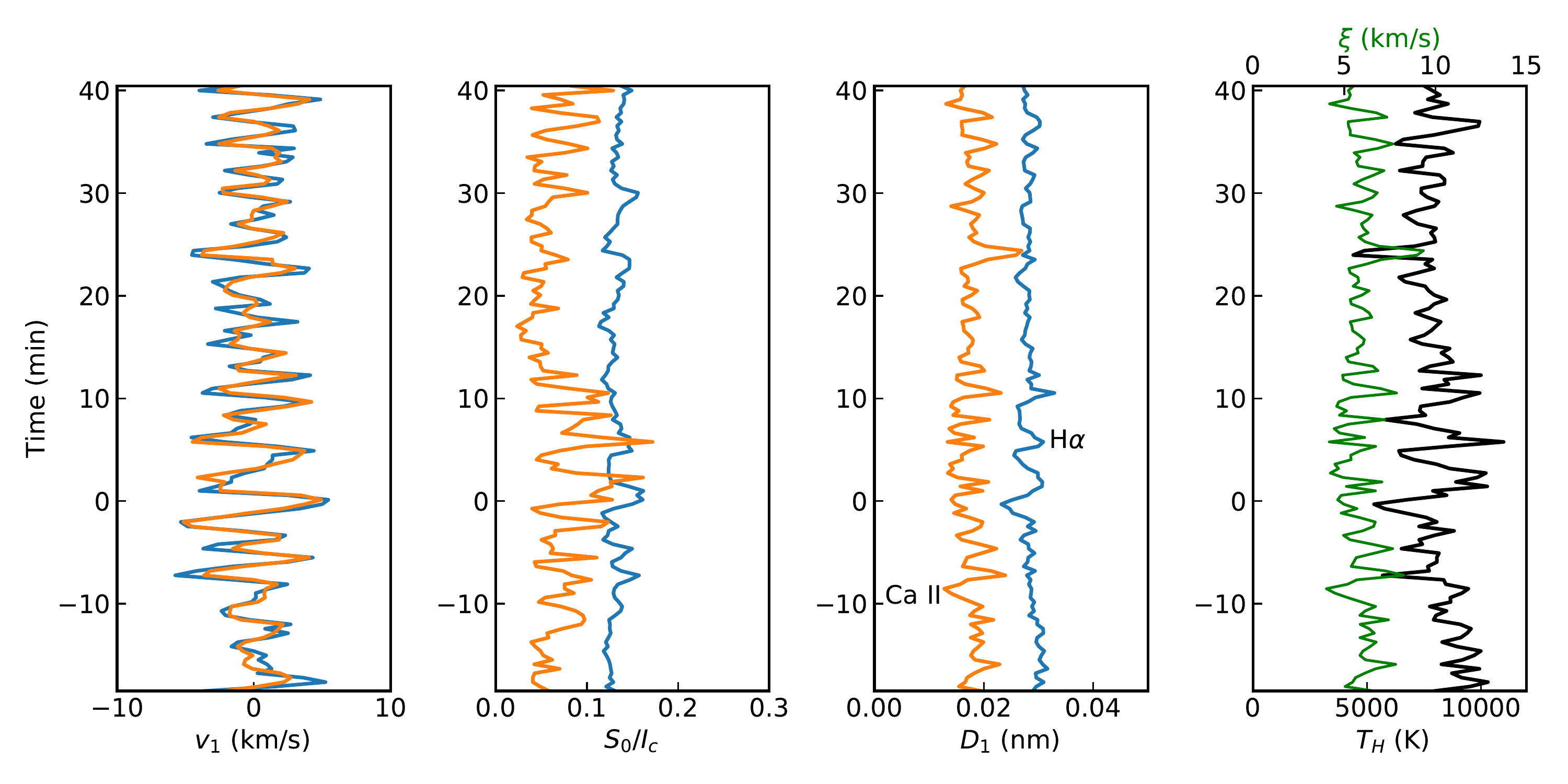}
  \caption{Temporal variations of some of the model parameters as well as $T_{\rm H}$ and $\xi$. }\label{timepar}
\end{figure*}

We confirm from \fg{sfT} that $S_0$ of the \ca\ line  is strongly correlated with $T_{\rm H}$.  The Pearson correlation coefficient is  0.75.  This means that the value of  $S_0$ , or equivalently, the  core intensity, is sensitive to temperature. This result supports the study of \cite{Cauz2009}. The temperature sensitivity of the core intensity of \ca\ line  originates from the property of the \ca\ line. In this line, the collisional excitation  by electrons significantly contributes to the \ca\ line emission. In contrast, the \ha\ line $S_0$ or the core intensity is not correlated with $T_{\rm H}$ at all, in agreement with \cite{Leen12}, which means that the collisional excitation by electrons is less important in the \ha\ line emission. The \ha\ core intensity is correlated with the average formation height, with the lower intensity corresponding to the higher average formation height \citep{Leen12}.

\subsection{Temporal variations}

It is interesting to determine the variation in model parameters with time at a fixed point. \fg{wlt} presents the line profiles of each line that were taken from an IN point and  stacked over time (approximately one hour). The time epoch was set to the instant of the highest downflow speed.  We find from the figure that the observed data
  ($\lambda-t$ intensity maps) are fairly well reproduced by the corresponding  models. The mean value of the fitting error $\epsilon$ was estimated at  0.0062 in the \ha\ line and at 0.014 in the \ca\ line. The residual intensity maps show some systematic patterns that are correlated with the intensity distribution.  Systematic errors seem to contribute significantly to the fitting errors. Some vertical patterns  represent telluric  lines and weak photospheric lines. These patterns are outside the clean wavelengths and were excluded in the inversion.

\fg{timepar} clearly indicates that all the model parameters as well as $T_{\rm H}$  and $\xi$  fluctuate with time.   The standard deviations of the fluctuations are 2.4 \kms\ in the \ha\ line $v_1$,  2.0 \kms\ in the \ca\ line $v_1$,   1100 K in $T_{\rm H}$,  and 0.9 \kms\ in $\xi$.   These are far larger than the estimates of the random errors 0.06 \kms, 0.04 \kms, 140 K, and 0.1 \kms, respectively. Thus the fluctuations shown in \fg{timepar} may represent the real variations in the physical conditions.  The variations of $v_1$ of the two lines represent chromospheric three-minute oscillations.  The temporal variations  of $T_{\rm H}$  and $\xi$ may be mostly attributed to the three-minute oscillations as well.  The proper interpretation of the variations of $T_{\rm H}$ and $\xi$ in terms of  chromospheric oscillations, however, is not straightforward, and  requires further investigation to  understand how the three-minute oscillations affect the formation of the two lines and  their model parameters of the multiline spectral inversion.  The three-minute oscillations in the upper chromosphere are now understood as highly nonlinear waves  of very long wavelength  that have sharp discontinuities, for example, shock fronts \citep{2017ApJ...844..129C, 2018ApJ...854..127C}.

\section{Discussion}

We proposed a multilayer spectral inversion in order to infer the physical condition of chromospheric plasma from strong absorption lines, and we specifically investigated the three-layer model in detail.  This model is fully specified by 13 parameters. By fixing some parameters and determining one parameter from a weak line that is formed in the photosphere,  we reduced the number of free parameters to 8 and determined these free parameters by applying the constrained nonlinear least-squares fitting.
We found that the three-layer model reproduces  most of the observed profiles of the \ha\ line and the \ca\ 854.2 nm line taken from a quiet-Sun region fairly well. The random fitting errors are much smaller than the intrinsic spatial variations, therefore  the maps of most model parameters look very regular.  Thus, we conclude that our implementation of the spectral inversion is successful in determining the model parameters, and we have reached the first goal of successfully implementing the three-layer spectral inversion.

The ultimate goal of our investigation is to infer the physical conditions from the determined model parameters.  This goal is distinct from the first goal  and requires further investigation.
 The physical condition  at each time is to be specified by the height variations in temperature, electron density, velocity, and nonthermal speed, but the determined model parameters are only a limited number of height-averaged source function, Doppler width, and  line-of-sight velocity in each layer. Moreover, the spectral inversion does not provide any information on the  height range of each layer.  Hence, adopting some assumptions is indispensable to derive the  physical conditions from the model parameters. The physical plausibility of the derived physical condition depends on the validity of the adopted assumptions.

 For instance,  hydrogen temperature $T_{\rm H}$ and nonthermal speed $\xi$ of the upper chromosphere were derived from the model parameters:  $w_{\rm H}$  and $w_{\rm Ca}$, the Doppler widths  of the \ha\ and  the \ca\ 854.2 nm lines based on the assumption  that the upper chromosphere seen through the \ca\ line is the same as the upper chromosphere seen through  the \ha\ line.  Under this assumption,  the same values of $T_{\rm H}$ and $\xi$ contribute to both  $w_{\rm H}$ and $w_{\rm Ca}$.  In fact, $T_{\rm H}$ is mostly determined by $w_{\rm H}$  because a hydrogen atom is  lighter and has a higher thermal speed than any other atoms or ions, as was previously noted by \cite{Cauz2009} and \cite{Leen12}.

 We determined  $T_{\rm H}$ and $\xi$ in all the pixels of the observed quiet-Sun region and found that the distribution of the hydrogen temperature peaks at around 7000\,K in the intranetwork regions, and around 11000 K in network regions.  The mean value of nonthermal speed is found to be 6 \kms regardless of intranetwork regions and network regions. Our measurements may be compared with the estimates of \cite{Cauz2009}. They determined  $T_{\rm H}$ and $\xi$ in a quiet region from the core widths of the two lines.  Each core width was defined very like an FWHM and was directly measured from the line profile without taking the effect of radiative transfer into account. After subtracting  the  ``intrinsic contribution'' from the line widths that probably represents the opacity contribution to the line width,  they obtained   $T_{\rm H}$ ranging from 5000 K to 60000 K, and  $\xi$  ranging from 1 \kms\ to 11 \kms.   These ranges include and are broader than the corresponding ranges we obtained.
\cite{Cauz2009} also noted that the core width of the \ha\ line is  larger in the network patches, supporting the notion  that network regions are heated more strongly than intranetwork ones.

 The assumption used for the derivation of $T_{\rm H}$ and $\xi$ is reasonable, but it also has  certain limitations.  When we determine the volume-averaged values of $T_{\rm H}$ and $\xi$ in the upper chromosphere, the assumption seems to be good enough, as described above. If we were to  determine the values in a fine \ha\ structure such as a fibril, the assumption is not satisfactory.  The \ha\ fibrils are often invisible in the \ca\ 854.2 nm line, which means that these plasma structures are transparent in the \ca\ line.  This may partly explain the lack of regional dependence on $\xi$ we found above.   $\xi$ is mostly determined by the Doppler width of the \ca\ line, but fibrils  are transparent in the \ca\ line, whereas they are clearly visible in the \ha\ line. If fibrils have  higher temperature and higher nonthermal speed than the low-lying layers where the \ca\ line is formed,  it is likely  that $\xi$ in network regions   was underestimated, and $T_{\rm H}$ was overestimated.

 Making better use of the determined model parameters requires a good understanding of the line formation.  In this regard, one can learn much from the forward modelling of the non-LTE radiative transfer.  The forward modelling is different from the spectral inversion in that it initially adopts the height variation of temperature,  density, and velocity, and  calculates the source function by solving the rate equations for level populations and can determine the formation height.
 The core intensity of a line is close to the source function in the outermost part of  the line formation region. It is well known that the \ha\ source function in the outer layers is mostly determined by the radiation field and  is not sensitive to local temperature, but sensitive to height \citep{Leen12}. In contrast,  the \ca\ line source function in the outer layers is still affected by the collisional excitation and  is sensitive to the local electron temperature.  This explains  the strong correlation shown in \fg{sfT} and the similarly strong correlation between the \ha\ core width and the \ca\ 854.2 nm  core intensity reported by \cite{Cauz2009}.

 Finally,  we would like to mention that the effect of isotopic splitting  may have to be investigated  in the multilayer spectral inversion of the \ca\ 854.2 line in the future. \cite{2014ApJ...784L..17L} investigated this effect on the bisector and inversions. They showed that the line core asymmetry and inverse C-shape of the bisector of the \ca\ 854.2 nm line can be explained by the isotopic splitting: the larger difference in line-of-sight velocity difference  of more than 2 \kms\ can result from this.

We conclude that the multilayer spectral inversion  successfully infers the model parameters from the observed profiles of strong absorption lines. The model parameters can be used to derive the physical parameters of chromospheric plasma such as the temperature, when physically plausible assumptions are made and if the line formation  is well understood.  Determining subtle variations of physical parameters in space or in time requires further careful  investigations.  Combining the model parameters of several lines would be of much help in determining the height variation of physical parameters. We expect  that the multiline multilayer spectral inversion will serve as a powerful tool to infer the physical parameters of chromospheric plasma from observations.

\begin{acknowledgements}
JC greatly appreciates Juhyung Kang's assistance in the implementation of the method using the Python software.
This research was supported  by the National Research Foundation of the Korea (NRF-2019H1D3A2A01099143,  NRF-2020R1A2C2004616), and by the Korea Astronomy and Space Science Institute under the R\&D program(Project No. 2020-1-850-07) supervised by the Ministry of Science and ICT.
\end{acknowledgements}

\bibliographystyle{aa} 
\bibliography{export-bibtex}
\end{document}